# Digital resonance tuning of high-$Q/V_m$ silicon photonic crystal nanocavities by atomic layer deposition

Xiaodong Yang [1,2], Charlton J. Chen[1], Chad A. Husko, and Chee Wei Wong[2]

Optical Nanostructures Laboratory, Columbia University, New York, NY 10027

We propose and demonstrate the digital resonance tuning of high-$Q/V_m$ silicon photonic crystal nanocavities using a self-limiting atomic layer deposition technique. Control of resonances in discrete steps of 122 ± 18 pm per hafnium oxide atomic layer is achieved through this post-fabrication process, nearly linear over a full 17 nm tuning range. The cavity $Q$ is maintained in this perturbative process, and can reach up to its initial values of 49,000 or more. Our results are highly controllable, applicable to many material systems, and particularly critical to matching resonances and transitions involving mesoscopic optical cavities.

---

[1] These authors contributed equally to this work.

[2] Electronic mail: xy2103@columbia.edu, cww2104@columbia.edu

Two-dimensional photonic crystal (2D PhC) slabs confine light by Bragg reflection in-plane and total internal reflection in the third dimension. Introduction of point and line defects into 2D PhC slabs create localized resonant cavities and PhC waveguides respectively, with full control of dispersion ab initio. Such defect cavities in high-index contrast materials, such as monolithic silicon, possess strong confinement with subwavelength modal volumes ($V_m$) at ~ $(\lambda/n)^3$, corresponding to high field intensities per photon for increased nonlinear and nonclassical interactions. Moreover, photonic crystal cavities with remarkable high quality factors ($Q$) [1, 2] have been achieved recently, now permitting nanosecond photon lifetimes for enhanced light-matter interactions. The strong localization and long photon lifetimes in these high-$Q/V_m$ photonic crystal nanocavities are strong candidates for enhanced nonlinear optical physics, such as optical bistability [3-5] and Raman lasing [6, 7], and cavity quantum electrodynamics [8].

These applications require precise control of cavity resonances to achieve tuned spectral overlap between the cavity modes and the gain or emitter material for controlled light-matter interactions. The cavity resonances is strongly dependent on the fabricated lattice constant $a$ and the hole radius $r$ of photonic crystals; slight differences in the photonic crystal geometries will result in large differences (a few to tens of nanometers in wavelength) in the dispersion characteristics. Active tuning using thermal (with associated phonon broadening of quantum dots) [9] or piezoelectric effects [10] can be employed. A passive post-fabrication tuning mechanism is particularly demanded, without external input power, to precisely align the designed resonant wavelengths. Specifically, wet chemical digital etching techniques [11] were recently developed for GaAs photonic crystal nanocavities, where the controlled blue shift of the cavity resonance was around 2-3 nm/cycle. Additionally, condensation of Xe [12] or self-assembled monolayers (such as a 2-nm polypeptide monolayer) [13] can be used, where a 3-5 nm cavity red shift per monolayer was observed for the latter. To achieve hundreds of picometer cavity resonance tuning, thin films below one nanometer is needed to be etched or deposited.

Atomic layer deposition (ALD) is widely used for gate dielectric and capacitance memory applications due to its high dielectric constant, precise thickness control, and highly conformal properties. Several metal oxides such as aluminum oxide ($Al_2O_3$), hafnium oxide ($HfO_2$), and titanium dioxide ($TiO_2$) have been used in low temperature ALD thin film growth. All these materials have been widely used in optical coating applications with relatively high refractive indices (at 1.55 μm, $n$ = 1.88 for $HfO_2$ [14], $n$ = 1.57 for $Al_2O_3$ [15], and $n$ = 2.18 for $TiO_2$ [16]) and a wide band gap with low absorption from the near-ultraviolet to the mid-infrared. Recently, ALD has become a promising tool for the fabrication of high quality three-dimensional photonic crystals from inorganic (opal) and organic (patterned polymer) templates [17-20]. Photonic band structure tuning in a 2D periodic lattice was also demonstrated, with 12% tuning range and 0.005% precision based on a deposition rate of 0.51 Å $TiO_2$ per ALD cycle [21].

In this Letter, we report the post-fabrication digital resonance tuning of high-$Q/V_m$ silicon photonic crystal nanocavities using self-limiting atomic layer deposition of $HfO_2$. The vertical radiation from the top of nanocavities was collected to analyze the mode resonances after each conformal coating step with slightly decreased $r/a$ and increased $t/a$ ratios in air-bridged photonic crystal slabs. The results demonstrate wide tuning range and precise fine control of cavity resonances while preserving high quality factors. The observed deposition rate is around 0.93 Å $HfO_2$ per ALD cycle, which leads to the red shift of resonant wavelength with precision of 122 ± 18 pm for a resonant wavelength ~ 1.55 μm. Total resonant wavelength tuning range is around 17 nm and currently only limited by the number of deposition steps (seven; 140 ALD cycles) used in this study.

The structure investigated is an air-bridged triangular lattice photonic crystal slab with silicon membrane thickness of 190 nm ($t/a$ = 0.4524) and air holes radii of 90 nm ($r/a$ = 0.2143), where the lattice period $a$ = 420 nm. High-$Q/V_m$ nanocavities with five linearly aligned missing air holes (*L5*) are side coupled with photonic crystal waveguides, as shown in Figure 1(a). The shift $S_1$ of two air-holes at cavity edge is 0.02$a$, 0.06$a$ and 0.10$a$, respectively, for three different *L5* nanocavities studied, in order to tune the radiation mode pattern for increasing the $Q$ factors. The waveguide-to-cavity separation is five layers of holes. The devices were initially patterned with deep UV lithography at the Institute of Microelectronics in Singapore, and subsequently etched with $SF_6/C_4F_8$-based inductively coupled plasma (ICP) into the silicon-on-insulator substrate. Next, optical lithography with AZ4620 photoresist was used to open a window in photonic crystal region, and 10 minutes HF BOE (6:1) was used to release the air-bridged structures. Samples were then cleaned using Piranha ($H_2SO_4:H_2O_2$ 3:1) solution for 5 minutes followed by HF BOE (6:1) solution dip for 30 seconds and deionized water rinse. Hot methanol was used as the final rinsing liquid to prevent stiction due to its lower surface tension. This procedure results in ~ 6 Å of (O-H)-terminated silicon oxide on the surface of silicon air-bridged photonic crystal slabs [22]. All samples were exposed to UV generated ozone for 10 minutes to restore the hydrophilic character of surface immediately prior to $HfO_2$ deposition. Figure 1(a) shows the top-view scanning electron microscopy (SEM) image of air-bridged $L5$ cavity with $S_1 = 0.02a$ before ALD. Figure 1(e) shows the electric field $E_y$ of the resonance mode mid-slab from 3D finite-difference time-domain simulations, calculated using a freely available software package with subpixel smoothing for increased accuracy [23].

Thin films of amorphous $HfO_2$ are deposited conformally on silicon air-bridged photonic crystal slabs by means of ALD at 150°C. Films were deposited using tetrakis(diethylamido)hafnium (IV) [$Hf(DEA)_4$] and water ($H_2O$) vapor in alternating pulses with $N_2$ purge of the reaction chamber between pulses. Each deposition step includes 20 ALD cycles, and each cycle consists of $Hf(DEA)_4$ injection for 0.25 seconds, $N_2$ purge for 150 seconds, $H_2O$ injection for 0.02 seconds, and $N_2$ purge for 200 seconds. The observed linear deposition rate is around 0.93 Å per cycle, which is about a monolayer of hafnium oxide. We note that lower substrate temperature down to 90°C is possible with our machine, at the expense of slow deposition rates. Figure 1(b) shows the top-view SEM image of $L5$ cavity after seven deposition steps, with the same magnification as in Figure 1(a). Based on geometrical statistical analysis of high-resolution SEM images, the hole radius reduces from 92.84 ± 1.56 nm to 79.86 ± 2.66 nm [24]. Figure 1(c) and 1(d) are the angled-view SEM images of air-bridged photonic crystal slabs before ALD and after seven deposition steps, respectively. The surface is still smooth enough to support high-$Q$ modes for $L5$ nanocavities after $HfO_2$ deposition. The thickness of photonic crystal slabs increases from 190 nm to 216 nm based on SEM estimates. These geometry changes agree well with the deposition schematic cross section of the sample morphology drawn in Figure 1(f).

With slightly decreased *r/a* and increased *t/a* ratios in air-bridged photonic crystal slabs, the photonic band gap will shift to lower frequencies. In addition to a frequency shift, the photonic bandgap also decreases from an 11.4% to a 9.7% gap with a deposition of $HfO_2$, computed using a freely available software package [25]. This can be attributed to a lower-index contrast between the holes and the bulk dielectric. The resonant wavelength of *L5* nanocavities will undergo a red shift.

For the measurement setup, a polarization controller and a lensed fiber are used to couple transverse-electric (TE) polarization light from tunable laser source (1480-1580 nm, wavelength accuracy 10 pm with 200 kHz linewidth) into the waveguide. A second lensed fiber collects the transmission from the waveguide output to check the total transmission loss of the whole system, which is around 24.8 dBm at wavelength of 1550 nm. The vertical radiation from the top of nanocavities collected by a 40X objective lens (NA of 0.65) and a 4X telescope was sent to the photodetector and lock-in amplifier to analyze the cavity resonances. In order to exclude optical nonlinear effects, low input power of 10 μW was coupled to the waveguide. Figure 2(a) plots the measured cavity resonances after each deposition step for *L5* cavity with $S_1 = 0.02a$. Figure 2(b) magnifies the resonance peak after the fourth deposition step. The quality factor *Q* is estimated from the full-width at half maximum and is ~ 49,000. From the 3D FDTD method, the *Q* factor and modal volume are calculated around 50,000 and ~ 0.98 cubic wavelengths (($\lambda/n)^3$) respectively.

Figure 3(a) shows the tuned resonant wavelength scales linearly with the number of deposition steps for all three *L5* cavities under investigation. Total resonant wavelength tuning range is around 17 nm with the current seven deposition steps. With more deposition steps, wider tuning range can be obtained. The 3D FDTD simulations (inset of Figure 3(a)) show a linear shift in the resonant wavelength as expected from small perturbations, although there is more uncertainty in the simulations due to the high spatial resolution (~ a few nanometers or less) required to capture this digital tuning. Figure 3(b) plots the resonant wavelength increment for each deposition step. An average wavelength red shift of 2.443 ± 0.359 nm is obtained for each step, which corresponds to a resonance shift of 122 ± 18 pm per $HfO_2$ monolayer deposition. An oscillatory variation of the resonance shift is also observed, as shown in Figure 3(b). This could be due to variations in the film deposition thickness, which is not perfectly the same in each step. In addition, we observe that the resonance increment itself increases slightly from 2.2 nm to 2.7 nm based on the linear curve fit. This is because, due to the conformality of ALD process, more dielectric material will be added relative to the previous step due to the expanded surface area, so that the resonance increment also slightly goes up, as illustrated in deposition schematics in Figure 1(f).

With different deposition material, the precision of resonant wavelength shift per ALD cycle can be changed. Single monolayer of $HfO_2$ induces an average 122 pm shift ($n$ = 1.88 at 1.55 μm, 0.93 Å per ALD cycle at 150 °C). From first-order perturbation estimate, a monolayer of $TiO_2$ ($n$ = 2.18 at 1.55 μm, 0.5 Å per ALD cycle at 100 °C [16]) can induce approximately 54 pm shift, while a monolayer of $Al_2O_3$ ($n$ = 1.57 at 1.55 μm, 1 Å per ALD cycle at 100 °C [15]) can generate approximately a 158 pm wavelength shift.

Figure 3(c) illustrates the variation of quality factor $Q$ with the number of deposition steps for all three $L5$ cavities. After first deposition step, $Q$ values drop almost by half for all cavities. This is because the ALD deposited film has a larger roughness initially, leading to more surface and air hole sidewall roughness scattering. With subsequent deposition steps, the conformal deposition gives a smoother film surface, permitting the $Q$ values to recover back to almost their initial values. The $Q$ values always maintain at least 20,000 or more during the deposition steps; this characteristic is also observed in our 3D FDTD simulations. This demonstrated shift in the resonance, while preserving the cavity $Q$, in response to a monolayer deposition also suggests these cavities as possible integrated sensors with pronounced responsivity to environmental conditions.

In summary, we have developed a technique for fine tuning the resonant wavelengths of high-$Q/V_m$ silicon photonic crystal nanocavities digitally using ALD of $HfO_2$ monolayers. The results demonstrate a nearly linear tuning across a range of around 17 nm. The tuning range is currently limited only by the number of deposition steps used in this study. The tuning precision is 122 ± 18 pm per ALD cycle while preserving high quality factors of resonant modes in $L5$ photonic crystal nanocavities. With selective patterning, $HfO_2$ monolayers can be selectively deposited only within the nanocavity region using low-temperature ALD [26] for even finer tuning control. The highly controlled, digital tuning of high-$Q$ modes in silicon photonic crystal nanocavities allows for practical realization of optical devices involving multiple resonances and matching transitions between quantum dots and optical resonances for cavity quantum electrodynamics.

This work was partially supported by DARPA, the New York State Foundation for Science, Technology and Innovation, and the National Science Foundation (ECS-0622069). The authors acknowledge the support of Mingbin Yu, and Dim-Lee Kwong of the Institute of Microelectronics in Singapore. Xiaodong Yang acknowledges the support of an Intel Fellowship.

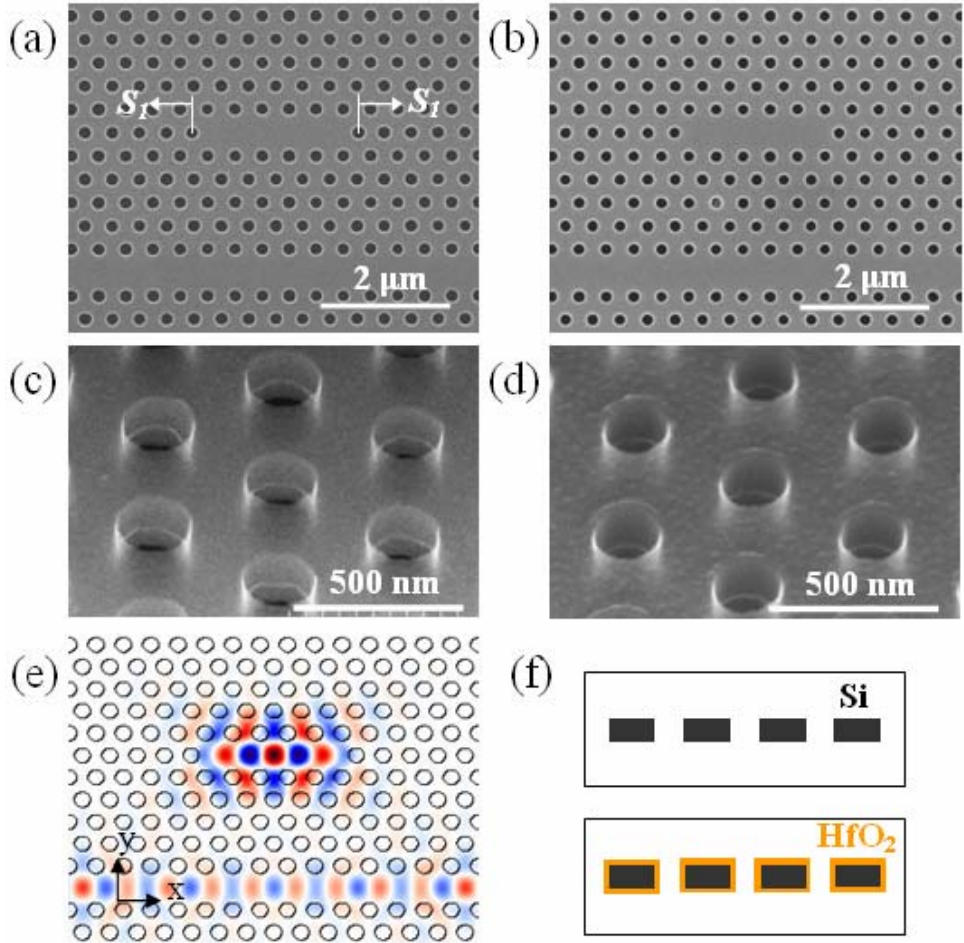

Fig. 1. (color online) Top-view SEM images of airbridge *L5* nanocavity with $S_1 = 0.02a$ coupled with photonic crystal waveguides (a) before ALD and (b) after 140 ALD cycles of $HfO_2$. 45º-angle-view SEM images of airbridge photonic crystal slabs (c) before ALD and (d) after 140 ALD cycles of $HfO_2$. (e) 3D FDTD calculated electric field $E_y$ profile of the high-*Q* mode supported in *L5* nanocavities. (f) Deposition schematic cross section of the sample morphology.

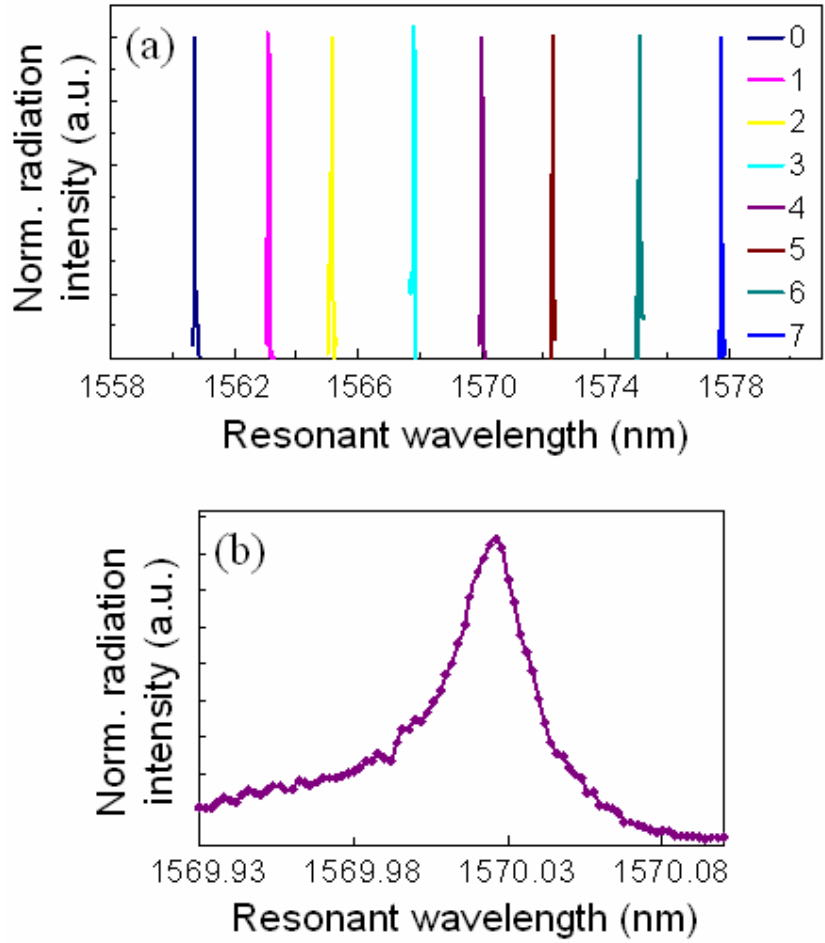

Fig. 2. (color online) (a) Measured cavity resonances after each deposition step ("1" to "7" in legend; "0" is unperturbed) for $L5$ cavity with $S_1 = 0.02a$. (b) Magnified resonance peak after the fourth deposition step, quality factor $Q \sim 49{,}000$.

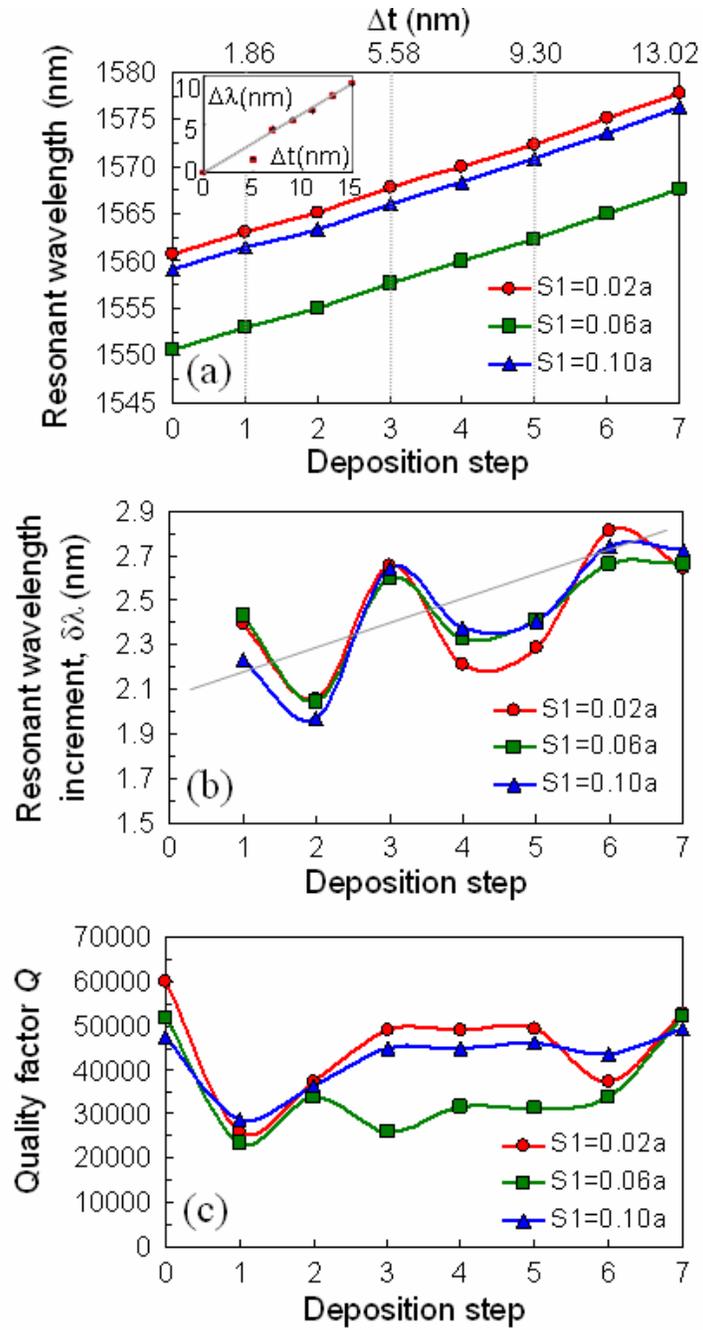

Fig. 3. (color online) (a) The tuned resonant wavelength scales linearly with the number of deposition step for all three *L5* cavities under investigation. Inset: 3D FDTD calculated wavelength shift (Δλ) for increasing thicknesses (Δt) of $HfO_2$ deposited, for all three cavities studied. (b) The wavelength increment (δλ) for each deposition step. (c) The variation of quality factor *Q* with the number of deposition steps for all three *L5* cavities.